\documentclass[aps,twocolumn,showpacs]{revtex4}
\usepackage{amssymb,graphicx}

\begin{document}

\title{Macroscopic Quantum Superposition and Entanglement in light reflection from
Bose-Einstein Condensates}
\author{Francesco S. Cataliotti$^{1}$ and Francesco De Martini$^{2,3}$}

\affiliation{(1) Dipartimento di Energetica and LENS, Universit\`{a} di Firenze, via N.
Carrara 1, I-50019 Sesto F.no (FI), Italy} 
\affiliation{(2) Dipartimento di Fisica, Universit\`{a} di Roma "La Sapienza", Piazzale
Aldo Moro 2, I-00185 Roma, Italy} 
\affiliation{(3) Accademia Nazionale dei
Lincei, Via della Lungara 10, I-00165 Roma, Italy.}

\begin{abstract}
The multiphoton quantum superposition generated by a quantum-injected
optical parametric amplifier (QI-OPA) seeded by a single-photon belonging to
an EPR entangled pair is made to interact with a \textit{Mirror-BEC} shaped
as a Bragg interference structure. The effect can be greatly enhanced if the
mirror is part of a cavity structure or is used within an optical
interferometer. The overall process will realize an Entangled Macroscopic
Quantum Superposition involving a ''microscopic'' singe-photon state of
polarization and the coherent ''macroscopic'' displacement of the BEC
structure acting in space-like separated distant places.
\end{abstract}

\maketitle

In recent years a great deal of interest has been attracted by the problem
of creating a Macroscopic Quantum Superposition of a massive object, e.g. a
tiny mirror, by an entangled interaction with a single photon, then
realizing a well known 1935 argument by Erwin Schroedinger \cite{Schroe35,
Marshall03, Bohm06}. The present work considers the adoption of the process
of Atom--Photon off resonant coherent scattering to create joint atom-photon
states nonseparably entangled by momentum conservation. The extension of
such property to macroscopic quantum states of light and matter would indeed
realize an Entangled Macroscopic Quantum Superposition \cite{Chan03}. Light
scattering from Bose--Einstein condensates have been used to enhance their
non--linear macroscopic properties in super--radiance experiments \cite
{Inouye99} and to show the possibility of matter wave amplification \cite
{Kozuma99} and non--linear wave mixing \cite{Deng99}. It has also been shown
that it is possible to transform a purely single particle phenomenon like
Rabi oscillations into a super--radiant Rayleigh scattering process which is
intrinsically macroscopic \cite{DeSarlo05}. Indeed Rayleigh scattering
experiments have demonstrated the reflection of light from bulk condensates\ 
\cite{CARL}. What we intend to discuss here is the linear \textit{coherent
scattering, e.g. }the\textit{\ reflection} by a Bose-Einstein condensate
(BEC) \cite{Inguscio} of a large assembly of nearly monochromatic photons
generated by a high-gain Optical Parametric Amplifier ''quantum-injected''\
in a nonlocal Einstein-Podolsky-Rosen (EPR) configuration, hereafter
referred to as QI-OPA \cite{DeMa98B, DeMa05, DeMa07}. To this aim we need to
structure a BEC in order to realize a high reflectivity mirror of photons,
the ''\textit{Mirror-BEC''}.

\begin{figure}[h]
\includegraphics[width=8 cm]{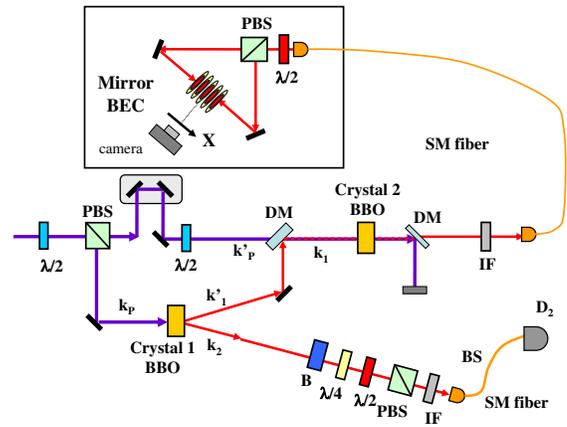}
\caption{Layout of the experimental apparatus}
\label{schema}
\end{figure}

Let's first account for the optical part of the apparatus: Figure \ref
{schema}. The adopted QI-OPA operates in the highly efficient collinear
regime and amplifies a single photon in a quantum superposition state of
polarization $(\overrightarrow{\pi })$, a \textit{qubit}, injected over the
spatial mode $\mathbf{k}$ into a large number of photons (in excess of $%
\overline{m}=10^{5}$) over the same mode $\mathbf{k}$ and with othogonal
polarizations. The injected photon belongs to a $\overrightarrow{\pi }-$%
entangled pair realized by an EPR\ optical configuration in two distant \
measurement stations, referred to as \textit{Alice} (A)\ and \textit{Bob}
(B)\ in the the jargon of quantum information. In virtue of the EPR\
nonlocality a measurement over the state of one photon by the Alice's
apparatus, results in a deterministic control at the Bob's site of the 
\textit{qubit,} injected in the QIOPA\ device. As investigated in previous
works, this device realizes the \textit{optimal} phase-covariant cloning of
injected qubits belonging to the equatorial circle of the corresponding
Bloch sphere which is orthogonal to the z-axis, the one that connects the
states of linear horizontal and vertical polarizations: $\left\vert
H\right\rangle $ and $\left\vert V\right\rangle $ \cite{Scia05, DeMa07}.
Accordingly, the information codified in the injected qubits $\left\vert
\varphi \right\rangle =2^{-1/2}(\left\vert H\right\rangle +e^{i\varphi
}\left\vert V\right\rangle )$, consists of the phase $\varphi $.

The QI-OPA\ device developed for the present experiment is a\ well tested,
working device whose properties are outlined here in some details \cite
{Naga07}. The excitation source (not shown in the Figure) is a Ti:Sa
Coherent MIRA mode-locked laser further amplified by a Ti:Sa regenerative
REGA device operating with pulse duration $180fs$ at a repetition rate of $%
100kHz$. The output beam, frequency-doubled by second-harmonic generation,
provides the excitation beam of UV wavelength (wl) $\lambda _{P}=397.5nm$.
The UV beam is split in two beams through a $\lambda /2$ waveplate and a
polarizing beam splitter $(PBS)$ and excites two BBO ($\beta $-barium
borate) nonlinear (NL) crystals cut for type II phase-matching. Crystal 1,
excited by the beam $\mathbf{k}_{P}$, is the spontaneous parametric
down-conversion (SPDC) source of entangled photon couples of wl $\lambda
=2\lambda _{P}$, emitted over the two output modes $\mathbf{k}_{i}$ ($i=1,2$%
) in the entangled \textit{singlet} state $\left| \Psi ^{-}\right\rangle
_{k1,k2}$= $2^{-1/2}\left( \left| H\right\rangle _{k1}\left|
V\right\rangle _{k2}-\left| V\right\rangle _{k1}\left| H\right\rangle
_{k2}\right) \ \label{SPDCentangled}$. This EPR device may be thought as
amplifying the vacuum fields associated with the output modes. The photon
associated to mode $\mathbf{k}_{2}$ (the \textit{trigger} mode) is coupled
to a single mode (SM)\ fiber and filtered by a set of $\overrightarrow{\pi }%
- $analysing optical devices, namely a Babinet compensator (B) a $\lambda /2$
$+$ $\lambda /4$ waveplate set, a (PBS)\ and an interference filter (IF)
with a\textit{\ hwhm} band $\Delta \lambda \approx 1.5nm$. At last the 
\textit{trigger} photon excites, at the Alice's site, the single photon
module (SPCM) $D^{A}_{2}$ , which delivers the the \textit{trigger }signal
adopted to establish the overall quantum correlations. By a Dichroic Mirror
(DM) the single photon generated over the mode $\mathbf{k}_{1}^{\prime }$is
made to merge into the mode $\mathbf{k}_{1}$ together with the UV pump beam
associatd with mode $\mathbf{k}_{P}^{\prime }$, and then injected into the
NL Crystal 2 where stimulates the emission of many photon pairs over the two
output $\overrightarrow{\pi }-$modes, $\mathbf{k}_{1}$. By virtue of the
nonlocal correlation acting on modes $\mathbf{k}_{1}^{\prime }$ and $\mathbf{%
k}_{2}$, the injected qubit is prepared in the state $\left| \varphi
\right\rangle _{k1}=2^{-1/2}(\left| H\right\rangle _{k1}+e^{i\varphi
}\left| V\right\rangle _{k1})$ by measuring the photon on mode $\mathbf{k}%
_{1}$ in the appropriate polarization basis, as said. The exact time and
space overlapping \ in Crystal 2 of the pump and injected pulses is assured
by means of an adjustable spatial delay $(Z)$.

The Crystal 2 is oriented for collinear operation over the two linear
polarization modes, H and V. The OPA\ interaction Hamiltonian $\widehat{H}%
=i\chi \hbar \left( \widehat{a}_{H}^{\dagger }\widehat{a}_{V}^{\dagger
}\right) +h.c.$ acts on the single spatial mode $\mathbf{k}_{1}$ where $%
\widehat{a}_{\pi }^{\dagger }$ is the one photon creation operator
associated with $\mathbf{k}_{1}$ and $\overrightarrow{\pi }$. The main
feature of this Hamiltonian is its \textit{phase-covariance}, i.e.
invariance under $U(1)$ transformations for qubits with equatorial
polarization leading for them to the \textit{optimality} of the cloning
process, as said. Owing to this property we can then re-write: $\widehat{H}$
=$\frac{1}{2}i\chi \hbar e^{-i\varphi }\left( \widehat{a}_{\pi }^{\dagger
2}-e^{i2\varphi }\widehat{a}_{\pi \perp }^{\dagger 2}\right) +h.c.$ for $%
\varphi \in (0,2\pi )$ where $\widehat{a}_{\pi }^{\dagger }=2^{-1/2}(%
\widehat{a}_{H}^{\dagger }+e^{i\varphi }\widehat{a}_{V}^{\dagger })$ and $%
\widehat{a}_{\pi \perp }^{\dagger }=2^{-1/2}(-e^{-i\varphi }\widehat{a}%
_{H}^{\dagger }+\widehat{a}_{V}^{\dagger })$. In particular we consider the
action over the states $\left\{ \left| +\right\rangle ,\left| -\right\rangle
\right\} $, e.g. with $\overrightarrow{\pi }_{\pm }=2^{-1/2}(%
\overrightarrow{\pi }_{H}\pm \overrightarrow{\pi }_{V})$ belonging to the
equatorial plane orthogonal to the z-axis. \ The injected state on mode $%
\mathbf{k}_{1}$ evolves into the output state $\left| \Phi \right\rangle
_{k1}=\widehat{U}\left| \varphi \right\rangle _{k1}$ according to the OPA\
unitary $\widehat{U}=\exp \left[ -i\widehat{H}t/\hbar \right] $, being $t$
the interaction time \cite{Pell03}. The overall amplified output state on
the $\mathbf{k}_{1}$spatial mode is expressed as: 
\begin{equation}
\left| \Phi \right\rangle ^{\pm }=\sum\limits_{i,j=0}^{\infty }\gamma _{ij}%
\frac{\sqrt{(1+2i)!(2j)!}}{i!j!}\left| 2i+1\right\rangle _{\pm }\left|
2j\right\rangle _{\mp }
\end{equation}
and $\gamma _{ij}\equiv C^{-2}(-\frac{\Gamma }{2})^{i}\frac{\Gamma }{2}^{j}$%
, $C\equiv \cosh g$, $\Gamma \equiv \tanh g$, being $g$\ the NL\ gain\emph{\ 
} \cite{Naga07}. There $\left| p\right\rangle _{+}\left| q\right\rangle _{-}$
stands for a state with $p$ photons with polarization $\overrightarrow{\pi }%
_{+}$ and $q$ photons with $\overrightarrow{\pi }_{-}.$ The Macro-states $%
\left| \Phi \right\rangle ^{+}$, $\left| \Phi \right\rangle ^{-}$are
orthonormal, i.e.$^{i}\left\langle \Phi |\Phi \right\rangle ^{j}=\delta _{ij}
$. Note that the entangled state $\left| \Sigma \right\rangle _{k1,k2}=$\ $%
2^{-1/2}\left( \left| \Phi \right\rangle _{k1}^{+}\left| +\right\rangle
_{k2}-\left| \Phi \right\rangle _{k1}^{-}\left| -\right\rangle _{k2}\right) $%
, keeps its \textit{singlet} character in the multi-particle regime, and
expresses the nonlocal correlations between two distant objects: the \textit{%
Microscopic }(single particle) system expressed by the \textit{trigger}
state (mode $\mathbf{k}_{2}$) and the \textit{Macroscopic (}multiparticle)
system ($\mathbf{k}_{1}$) \ It is referred to in the literature as the ''%
\textit{Schroedinger Cat State''\ }[cfr\textit{\ }Schleich\cite{Schl01},
Ch.11, pgs. 306-319]. At the output of crystal $2$ the output beam with with
wl $\lambda \;$is spatially separated by the pump UV beam by a DM and an
interferential filter ($IF$) with bandwidth $1.5nm$ \textit{hwhm }and
finally coupled to a single mode optical fiber $(SM)$. The corresponding
output signals registered by any detector coupled to mode $\mathbf{k}_{1}$
are taken in coincidence with a corresponding single-photon signals which,
registered by the detector $D_{2}$ on the Alice's site, triggers nonlocally
the QI-OPA\ dynamics. Then we deal here with a ''conditional experiment''\
by which each significant event registered at the output of the QI-OPA\ is
conditioned by the actual realization of an EPR pair by the crystal 1 and
then by the actual ''quantum injection''\ of the crystal 2.\ This removes
efficiently the noise due to the SPDC in crystal 2.\ Let us now analyze the
output field $\mathbf{k}_{1}$ over the polarization modes $\overrightarrow{%
\pi }_{\pm }$ when the qubit $\left| \varphi \right\rangle _{k1}$is
injected. The ensemble average photon number $N_{\pm }$ over $\mathbf{k}_{1}$
with $\overrightarrow{\pi }_{\pm }$ is easily evaluated on the basis of \
Eq. 1 and found to depend on the phase $\varphi $ as follows: $N_{\pm
}(\varphi )\;$= $^{\pm }\left\langle \Phi \right| \widehat{N}_{\pm }(\varphi
)\left| \Phi \right\rangle ^{\pm }$ = $\overline{m}+\frac{1}{2}(2\overline{m}%
+1)(1\pm \cos \varphi )$ with $\overline{m}=\sinh ^{2}g$, the average value
of the number of \ ''squeezed vacuum''\ photons emitted by the OPA\ for each
polarization mode in absence of quantum injection \cite{DeMa98B}. In a
actual laboratory test a value of the gain in excess of $g=6$ could be
easily obtained. This corresponds to an average number of QI-OPA generated
photons per mode in excess of: $\overline{m}=$\ $80.000$. The photon number
difference: $N(\varphi )\;$= $[N_{+}(\varphi )-N_{-}(\varphi )]$ = $[(2%
\overline{m}+1)\cos \varphi ]\;$can give rise, as we shall see at once, to an%
\textit{\ }interference (IF)\ fringing pattern as function of the phase $%
\varphi $ of both the quantum-injected qubit and of the\ correlated, far
apart ''trigger''\ qubit. In other words, in virtue of the EPR\ nonlocality,
the quantum superposition implied by the single-particle\ ''trigger''\ qubit
measured by Alice in the \textit{microscopic} domain can manifest itself at
Bob's site in the \textit{macroscopic} (multiparticle) regime by a\ \textit{%
first-order coherence} interference (IF) fringing pattern \cite{Lou83}.\ It
has been tested experimentally that this interference behavior is
independent of the equatorial orthogonal basis chosen to represent the input
qubit and that the QI-OPA\ realizes a genuine quantum superposition rather
than a mixture \cite{DeMa07}. The\ IF\ fringe ''\textit{visibility}''\ $%
\mathcal{V}^{(1)}$=$(N_{\max }-N_{\min })/(N_{\max }+N_{\min })$ is
dependent on the gain $g\;$as follows:$\;\mathcal{V}_{theor}^{(1)}$= $(2%
\overline{m}+1)/(4\overline{m}+1)$. For $M\rightarrow \infty $, viz. $%
g\rightarrow \infty $, $\mathcal{V}_{th}^{(1)}\mathcal{\ }$of this
first-order correlation function attains the asymptotic value = $1/2$ 
\cite{Lou83}.

\begin{figure}[h!]
\includegraphics[width=5 cm]{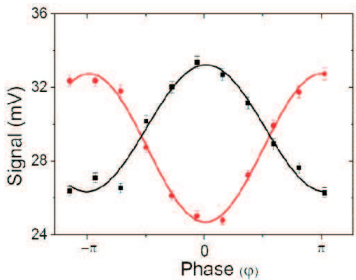}
\caption{Fringe patterns due to detection of the Macro-states $%
|\Phi^+\rangle $(red) and $|\Phi^+\rangle$ (black) as function of the phase $%
\protect\phi$ of the trigger qubit}
\label{fringes}
\end{figure}

How to detect the Macroscopic quantum superposition? The method successfully
adopted in our Laboratory was to separate the two orthogonal polarizations$\;%
\overrightarrow{\pi }_{\pm }\;$by inserting a $(PBS)$ on the output mode $%
\mathbf{k}_{1}$. The corresponding photon numbers $N_{\pm }(\varphi )$ were
detected by a couple of \ Burle C31034-A02 Ga-As photomultipliers with $\eta
_{QE}=13\%$\ and the difference of the output currents: $I(\varphi
)=(I_{+}-I_{-})\varpropto N(\varphi )\;$was registered. Two independent IF
fringing patterns obtained by this technique for opposing phases $\varphi $
and $\varphi +\pi $ are shown in Figure 2 \ref{fringes}. \ That result may
considered the preparatory stage of the present more sophisticated
experiment, sketched in Figure 1.


At the output of the QI-OPA two counteracting optical beams associated with
the the Macro-states $\left| \Phi \right\rangle ^{\pm }$ and carrying
respectively an average of $N_{+}(\varphi )$ and $N_{-}(\varphi )$ photons,
for a total of $N(\varphi )\approx 10^{5}$ are spatially separated \ and
focused on the opposed sides of a cigar-shaped Bose Einstein Condensate
(BEC) with approximate diameter 10$\mu m$. Since a photon exchanges a
momentum $p=2\hbar \nu /c\;$upon a head on collision with an atom, the
counteracting momenta of the two beams can be very efficiently transferred
to the BEC if \ this one is conveniently shaped in order to operate as a
mechanical mirror, indeed a \textit{Mirror-BEC}.


%
%
%

Let us illustrate how we intend to realize such a mirror, if we load a BEC
from a magnetic trap into a 1--D optical lattice we get an array of disk
shaped condensates with a longitudinal size $R_{l}\propto s^{-1/4}k^{-1} $ 
\cite{nota} and with a transverse size $R_{\perp }$ dictated by the strength
of the magnetic trap and by the number of atoms $N_{at}$ in each condensate 
\cite{Pedri01}. The number $N_{D}$ of such disks is also fixed by the
strength of the magnetic trap (this time in the longitudinal direction) and
by the number of atoms in the original condensate. Typical numbers are $%
N_{D}=100\div 200$, $N_{at}=100\div 1000$ with $R_{\perp }=2\div 10\mu $m
and $R_{l}=80\div 300\;$nm \cite{Morsch06}.

By choosing $s$ it is then possible to prepare an array of disks with a
longitudinal size of $\frac{\lambda}{4}$ spaced by $\frac{\lambda}{2}$. If
we approximate each condensate with a slab of dielectric with index of
refraction $n_B$ then this configuration is the same as that of a Bragg
mirror for the wavelength $\lambda$.

\begin{figure}[h]
\includegraphics[width=3cm]{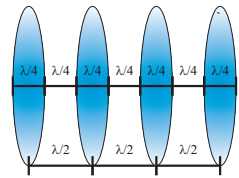}
\caption{Alternating slabs of condensate and vacuum.}
\label{slabs}
\end{figure}

The reflectivity of a Bragg mirror with $2N_{D}$ alternating layers of index
of refraction $n_{B}\sim (1+\epsilon )$ is: $R=\left( \frac{n_{B}^{2N_{D}}-1%
}{n_{B}^{2N_{D}}+1}\right) ^{2}\sim N_{D}^{2}\epsilon ^{2}$ with a
bandwidth: $\Delta \nu _{a}=\frac{4\nu }{\pi }\arcsin \left( \frac{n_{B}-1}{%
n_{B}+1}\right) \sim \frac{2\nu }{\pi }\epsilon $. For a 2--level atom very
far from resonance we can approximate $\epsilon \simeq \frac{3\pi }{2}{%
\mathcal{N}}\frac{\Gamma }{\Delta }$ where ${\mathcal{N}}=\lambdabar^{3}%
\frac{N}{V}$ is the rescaled density, $\Gamma $ is the atomic linewidth and $%
\Delta $ the detuning from resonance. In rubidium $\Gamma \simeq $ 6 MHz and
typical densities are $\frac{N}{V}=10^{13}\div 10^{14}$cm$^{-3}$. It should
be stressed that in the same approximation the light scattering probability
is $\sim N_{at}\frac{\Gamma }{\Delta }$. To be more precise the density is
given in the Thomas--Fermi approximation by $n_{TF}=\frac{1}{8\pi }\frac{1}{%
a_{ho}a}\left( \frac{15N_{at}a}{a_{ho}}\right) ^{\frac{2}{5}}$ where $a$%
(=5.77 nm in $^{87}$Rb) is the scattering length and $a_{ho}$ is the
averaged harmonic oscillator length. Combining all the previous equations
and assumptions we obtain the following graphics:

\begin{figure}[h!]
\includegraphics[width=6.5cm]{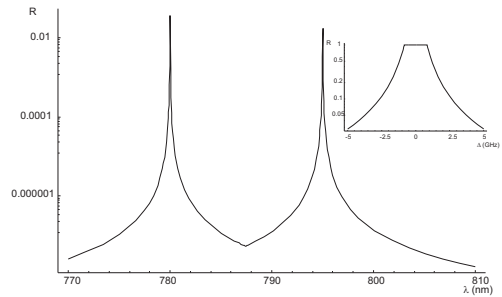}
\caption{Reflectivity of a patterned BEC as a function of wavelength. In the
inset is shown the reflectivity around resonance}
\label{Reflect}
\end{figure}

From the inset it is clear that around the atomic resonance with a bandwidth
of around \emph{\ }$\Delta \nu _{a}=2GHz$ the reflectivity of the patterned
BEC is essentially unity. Since the spectral width of the QI-OPA generated
photon beams is \emph{\ }$\Delta \lambda \sim 1.5nm$, corresponding to a
linewidth\emph{\ }$\Delta \nu \sim 700GHz$, about the 3\% of the \ incoming
photon beams will be totally reflected by the Mirror-BEC. In our experiment
this will corresponds to a number of active photons $N_{\pm }^{\prime
}(\varphi )=(\Delta \nu _{a}/\Delta \nu )\times N_{\pm }(\varphi )\;$in the
range $(150\div 300)$, having taken into account the limited value of the
visibility $\mathcal{V}^{(1)}\simeq 13\%$ measured experimentally at the
output of the QI-OPA. This effect is easily detectable on the basis of \ the
previous measurement on Rayleigh scattering from a BEC \cite{Cola04}. It
must be noted that very close to resonance all the above approximations
fail, however atomic resonance lines have linewidths of the order of a few
MHz which means that our pulses will contain around 1 photon at resonance
leaving a significant frequency band available for the proposed experiment.
A more careful model should take into account a quasi sinusoidal modulation
of the refractive index reflecting the atomic density distribution in the
Thomas--Fermi approximation. Such a model would result in a more physical
smoothing of the reflection coefficient around resonance. However the
simplified approach adopted here captures the main physical aspect \emph{i.e.%
} a significant enhancement of elastic light reflection around resonance. An
alternative approach would be to consider the second quantized model of
atom/photon fields interaction. Such a model goes far beyond the scope of
the present paper. Let us mention here that it has been introduced in the
framework of the ''Collective atomic recoil Laser'' \cite{CARL} and already
used to predict entanglement between atomic p-states \cite{Cola04}

It is important to note that, in order to observe recoil effects of the 
\emph{Mirror-BEC}, the condensate has to be released from the optical
lattice; indeed typical expansion velocities for a BEC are of the order of 1
nm/$\mu$s which leaves at least 50 $\mu$s before the pattern gets
significantly spoiled. Given the time duration of a QI-OPA pulse (order of 1
ps) there is enough time to observe a reflection from the free standing 
\emph{Mirror-BEC}. The release of the condensate is triggered by 
any UV pump laser pulse exciting the QI-OPA. 
During the free survival time of the the wavy structure of the Mirror-BEC, 
the two momentum exchanging $N_{\pm }(\varphi )\ $photon pulses emitted from
the QI-OPA\ are injected simultaneously from opposite sides in the BEC\
chamber, as said. By a fast electro-optic switch placed at the output of the
QI-OPA, this p-transfer process could be repeated by any large number of
times, say $10^{4}$. This device, recently developed in our Laboratory, is
deterministically driven by the measurement outcomes registered at the
Alice's station and\ then can enhance by a coherent add-on process the
displacement of the Mirror-BEC due to any elementary interaction:$\;\Delta
x(\varphi )$ $\varpropto $ $(2\hbar \nu /c)\times N_{\pm }^{\prime }(\varphi
)\;$= $[2\hbar \nu (2\overline{m}+1)/c]\cos \varphi $.$\;$Remind the
Micro-Macro entanglement structure of the state $\left| \Sigma \right\rangle
_{k1,k2}$. The realization of the interference fringe pattern $\Delta
x(\varphi )$ as function of the phase $\varphi $ of the single
photon-measured by the far apart Alice's apparatus will be a clear
demonstration of \ the entanglement between the Microscopic single-particle
realm and the mechanical motion of a Macroscopic, multi-atom BEC. Note that
the totally reflected photons bear the exact bandwidth for further total
reflection by the same Mirror-BEC. Thus the effect can be enhanced several
times by means of\ a couple of \ additional mirrors reflecting back the
optical beams to the same BEC after the first interaction. This idea bring
us straightforwardly to an electromagnetic \textit{cavity} structure by
which the BEC\ displacing effect can be enhanced by the cavity quality
factor $Q$, \emph{i.e.} in principle by orders of magnitude. An alternative
interesting picture would consider the \emph{Mirror-BEC} as the key part of
a Fabry-Perot or a Michelson interferometer.Furthermore the reflected
photons, trapped in any \emph{Mirror-BEC} cavity structure or optical
interferometer, will be themselves entangled with the condensate. This would
open a rich perspective of novel quantum physics enlightening the
nonlocality for Macroscopic systems, a fundamental long sought problem of \
modern Science \cite{Schroe35}.

We acknowledge interesting discussions with Fabio Sciarrino, Chiara Fort,
Leonardo Fallani, Massimo Inguscio. F.D.M. also acknowledges early
discussions with Markus Aspelmeyer leading to the basic idea of the present
proposal. The work is supported by MIUR (N. PRIN 06, PRIN 05) and CNISM\
(Progetto Innesco 2006).


\begin{thebibliography}{99}
\bibitem{nota}  $k$ is the wavenumber of the laser creating the optical
lattice; $s$ is the (adimensional) height of the optical potential in units
of the kinetic energy acquired by a condensate atom recoiling an optical
lattice photon.
\end{thebibliography}
\end{document}